\documentclass[10pt,letterpaper,twocolumn]{article} %% two column, final layout

\usepackage{ol}
\usepackage{hyperref}
\usepackage{amsmath}

\begin{document}

\twocolumn[ %% activate for two-column option

\title{Three-Port Beam Splitters/Combiners for Interferometer Applications}

\author{R. Schnabel, A. Bunkowski, O. Burmeister, and K. Danzmann}

\address{Max-Planck-Institut f\"ur Gravitationsphysik (Albert-Einstein-Institut), Universit\"at Hannover, \\Callinstr. 38, 30167 Hannover, Germany}

% Do not use \email or \homepage here. E-mail and URL can be given just before references.

\begin{abstract}
We derive generic phase and amplitude coupling relations for beam
splitters/combiners that couple a single port with three output
ports or input ports, respectively. We apply the coupling
relations to a reflection grating that serves as a coupler to a
single-ended Fabry-Perot ring cavity. In the impedance matched
case such an interferometer can act as an all-reflective ring
mode-cleaner. It is further shown that in the highly under-coupled
case almost complete separation of carrier power and phase signal
from a cavity strain can be achieved.
%One interferometer output port contains all the carrier power and the other port contains no carrier power but almost the complete signal.
\end{abstract}


 ] %% activate for two-column option

\noindent%\section{Introduction}

Two-port beam splitters/combiners, for example the partially transmitting mirror, are key devices in laser interferometry. They serve as 50/50 beam splitters in Michelson interferometers and as low transmission couplers to cavities. Amplitude and phase relations of two-port beam splitters/combiners are well-known.
In the case of grating optics, diffraction orders of a greater number can couple to one input port.
Recently, a reflection grating with three diffraction orders was
used for interferometer purposes; laser light was coupled into a
linear high finesse Fabry-Perot cavity using the second-order
Littrow configuration \cite{BBBDSCKT04}.
The grating was built from a binary structure. This property
together with the second-order Littrow configuration provided a
symmetry against the grating's normal. The system was theoretically
analyzed in \cite{BBDS05}. It was shown that the new three-port
(3p) coupled Fabry-Perot interferometer can be designed such that
resonating carrier light is completely back-reflected towards the
laser source. The additional interferometer port is then on a dark
fringe and contains half of the interferometer strain signal.

\begin{figure}[ht!]
    \centerline{\includegraphics[angle=0,width=6.2cm]{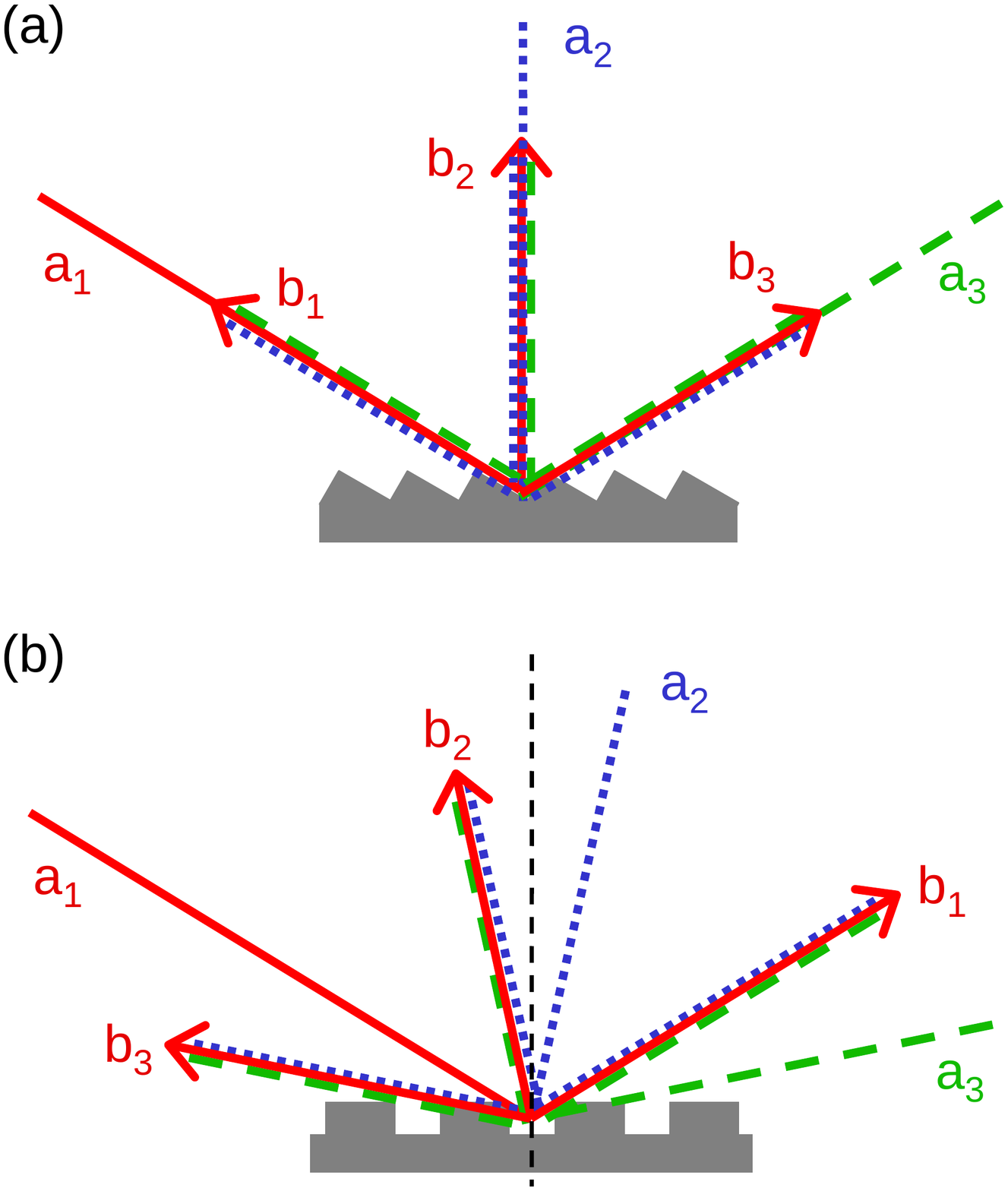}}
  \vspace{0mm}
  \caption{Two examples of three-port (3p) beam splitters/combiners. Input fields $a_i$ and output fields $b_i$ denote complex amplitudes of the electric field.
a)~Asymmetric triangular grating in second order Littrow
configuration. b)~Binary grating in non-Littrow configuration.
  }
  \label{3portgrat6}
\end{figure}

In this letter we first derive the generic coupling relations of
three-port (3p) beam splitters. This includes coupling
amplitudes as well as coupling phases which are required for
interferometric applications. Our description includes arbitrary gratings with three orders of diffraction
regardless of the groove shape and the diffraction angles, as
shown in Fig.~\ref{3portgrat6}. We then
investigate the three-port reflection grating coupled Fabry-Perot
ring interferometer and show that for a resonating carrier a dark
port can be constructed that contains an arbitrary high fraction
of the interferometer's strain signal.

%\section{Theory}

Optical devices can be described by a scattering matrix formalism \cite{Siegmann86}. In general the coupling of $n$ input and $n$ output ports require an $n \times n$ scattering matrix $\textbf{S}$. The $n$ complex amplitudes of incoming and outgoing fields are combined into vectors $\textbf{a}$ and $\textbf{b}$, respectively. For a loss-less device $\textbf{S}$ has to be unitary to preserve energy, and %for a symmetric labeling of the input and output ports as given in Figs.~\ref{3portgrat6},\ref{3portgrat6}
reciprocity demands $|S_{ij}|\equiv|S_{ji}|$
%$S_{ij}\equiv S_{ji}$
for all elements $S_{ij}$ of $\textbf{S}$. For a generic
three-port device 6 coupling amplitudes and 9
coupling phases are involved. Since 3 input and 3 output fields
are considered the number of phases can be reduced to 6
without loss of physical generality; the remaining 6 phases
describe the phases of the 6 fields with respect to a local
oscillator field. Here we choose the phases such that the matrix
$\textbf{S}$ is symmetric, and $\textbf{b} =
\textbf{S} \times \textbf{a}$ can therefore be written as
\begin{equation}
\label{Smatrix} \left( \begin{array}{c}
        b_{1}   \\
        b_{2}   \\
        b_{3}   \\
       \end{array}
\right )
 =
\left( \begin{array}{cll}
        \eta_{1} e^{i\phi_{1}}   &  \eta_{4} e^{i\phi_{4}}  &   \eta_{5} e^{i\phi_{5}}  \!\!\\
        \eta_{4} e^{i\phi_{4}}   &  \eta_{2} e^{i\phi_{2}}  &   \eta_{6} e^{i\phi_{6}} \!\!\\
        \eta_{5} e^{i\phi_{5}}   &  \eta_{6} e^{i\phi_{6}}  &   \eta_{3} e^{i\phi_{3}}  \!\!\\
       \end{array}
\right ) \!\times \! \left( \begin{array}{c}
        a_{1}   \\
        a_{2}   \\
        a_{3}   \\
       \end{array}
\right )\,,
\end{equation}
where $0 < \eta_{i} < 1$ for all $i$
describes the amplitude and $e^{i\phi_{i}}$ the phase of coupling.
Fig.~\ref{3portgrat6} shows two examples of three-port devices. In
both cases the input beam splits into three beams, and vice versa
three input beams can interfere into a single one.
%Both examples reveal a symmetry that justifies the symmetric form of the scattering matrix in Eq.~\ref{Smatrix}.
However, one realizes that the rigorously defined scattering
matrix for the device in Fig.~\ref{3portgrat6}b) has dimension $6
\times 6$; but this matrix contains null elements because not 6
but only 3 ports couple and the matrix can be reduced to the matrix as given in Eq.~(\ref{Smatrix}).

The unitarity condition $\textbf{S}^\dag \textbf{S} = \textbf{1}$
entails the following set of equations:
\begin{eqnarray}
\label{I}
1&\!\!=\!\!&\eta_{1}^2 + \eta_{4}^2 + \eta_{5}^2  \,,\;  \\[2mm]
\label{II}
1&\!\!=\!\!&\eta_{2}^2 + \eta_{4}^2 + \eta_{6}^2  \,,\;  \\[2mm]
\label{III}
1&\!\!=\!\!&\eta_{3}^2 + \eta_{5}^2 + \eta_{6}^2  \,,\; \\[2mm]
\label{IV}
|\mbox{cos}(2\phi_{4}\!-\!\phi_{1}\!-\!\phi_{2})|&\!\!=\!\!&  \frac{|\eta_5^2\eta_6^2-\eta_1^2\eta_4^2-\eta_2^2\eta_4^2|}{2\eta_4^2\eta_1\eta_2} \,,\;\\[1mm]
\label{V}
|\mbox{cos}(2\phi_{5}\!-\!\phi_{1}\!-\!\phi_{3})|&\!\!=\!\!&  \frac{|\eta_4^2\eta_6^2-\eta_1^2\eta_5^2-\eta_3^2\eta_5^2|}{2\eta_5^2\eta_1\eta_3} \,,\;\\[1mm]
\label{VI}
|\mbox{cos}(2\phi_{6}\!-\!\phi_{2}\!-\!\phi_{3})|&\!\!=\!\!&  \frac{|\eta_4^2\eta_5^2-\eta_2^2\eta_6^2-\eta_3^2\eta_6^2|}{2\eta_6^2\eta_2\eta_3}\,,\; \\[1mm]
\label{VII}
|\mbox{cos}(\phi_{6}\!+\!\phi_{4}\!-\!\phi_{5}\!-\!\phi_{2})|&\!\!=\!\!&  \frac{|\eta_1^2\eta_4^2-\eta_2^2\eta_4^2-\eta_5^2\eta_6^2|}{2\eta_2\eta_4\eta_5\eta_6}\,,\;\\[1mm]
\label{VIII}
|\mbox{cos}(\phi_{6}\!-\!\phi_{4}\!-\!\phi_{5}\!+\!\phi_{1})|&\!\!=\!\!&  \frac{|\eta_3^2\eta_5^2-\eta_1^2\eta_5^2-\eta_4^2\eta_6^2|}{2\eta_1\eta_4\eta_5\eta_6}\,,\;\;\\[1mm]
\label{IX}
|\mbox{cos}(\phi_{6}\!-\!\phi_{4}\!+\!\phi_{5}\!-\!\phi_{3})|&\!\!=\!\!&   \frac{|\eta_2^2\eta_6^2-\eta_4^2\eta_5^2-\eta_3^2\eta_6^2|}{2\eta_3\eta_4\eta_5\eta_6}\,.\;\;\;\;\;\;
\end{eqnarray}

Eqs.~(\ref{I})-(\ref{IX}) set boundaries for physically possible coupling amplitudes and phases of the generic loss-less 3p beam splitter/combiner. The first three equations represent the energy conservation law and arise from the diagonal elements of the unitarity condition.
The next six equations arise from the off-diagonal elements. They are already simplified to contain just a single
cosine term.
However, it can be easily deduced that up to three phases in the
scattering matrix $\textbf{S}$ can be chosen arbitrarily. In this
analysis we choose the phases $\phi_{1},\phi_{2},\phi_{3}$ to be zero. This is a
permitted choice without introducing any restriction on possible
coupling amplitudes. Then the phases of the scattering matrix can be written as
\begin{eqnarray}
\label{p7}\nonumber
\phi_{1}&=&\phi_{2}\;\;=\;\;\phi_{3}\;\;=\;\;0\,,\\ [2mm]\nonumber
\phi_{4}&=&  - \frac{1}{2} \mbox{arccos}\left(\frac{\eta_1^2\eta_4^2+\eta_2^2\eta_4^2-\eta_5^2\eta_6^2}{2\eta_4^2\eta_1\eta_2}\right) -\frac{\pi}{2}    \,,       \\ [1mm]
\phi_{5}&=&    \frac{1}{2} \mbox{arccos}\left(\frac{\eta_4^2\eta_6^2-\eta_1^2\eta_5^2-\eta_3^2\eta_5^2}{2\eta_5^2\eta_1\eta_3}\right)    \,,\\[1mm] \nonumber
\phi_{6}&=&    - \frac{1}{2} \mbox{arccos}\left(\frac{\eta_2^2\eta_6^2+\eta_3^2\eta_6^2-\eta_4^2\eta_5^2}{2\eta_6^2\eta_2\eta_3}\right) + \frac{\pi}{2} \,.
\end{eqnarray}

It is interesting to note that the coupling relations
restrict the possible values of  $\eta_i$. Let us assume, a free
choice of $\eta_{4}^2$ and $\eta_{6}^2$ is desired, which then
immediately determines $\eta_{2}^2$ according to Eq.~(\ref{II}).
Substituting $\eta_{1}$ and $\eta_{3}$ using Eqs.~(\ref{I}) and
(\ref{III}), Eqs.~(\ref{IV}) to (\ref{IX}) provide the following
pair of inequalities that restricts the values of $\eta_{5}$ and
thereby also the values of $\eta_{1}$ and  $\eta_{3}$:
\begin{eqnarray}
\label{inequality} \frac{\eta_{4} \eta_{6} \,(1 -
\eta_2)}{\eta_{4}^{2} + \eta_{6}^2}  \leq \eta_{5} \leq
\frac{\eta_{4} \eta_{6} \,(1 + \eta_2)}{\eta_{4}^{2} +
\eta_{6}^2}\;.
\end{eqnarray}

\begin{figure}[ht!]
\centerline{\includegraphics[angle=0,width=7.7cm]{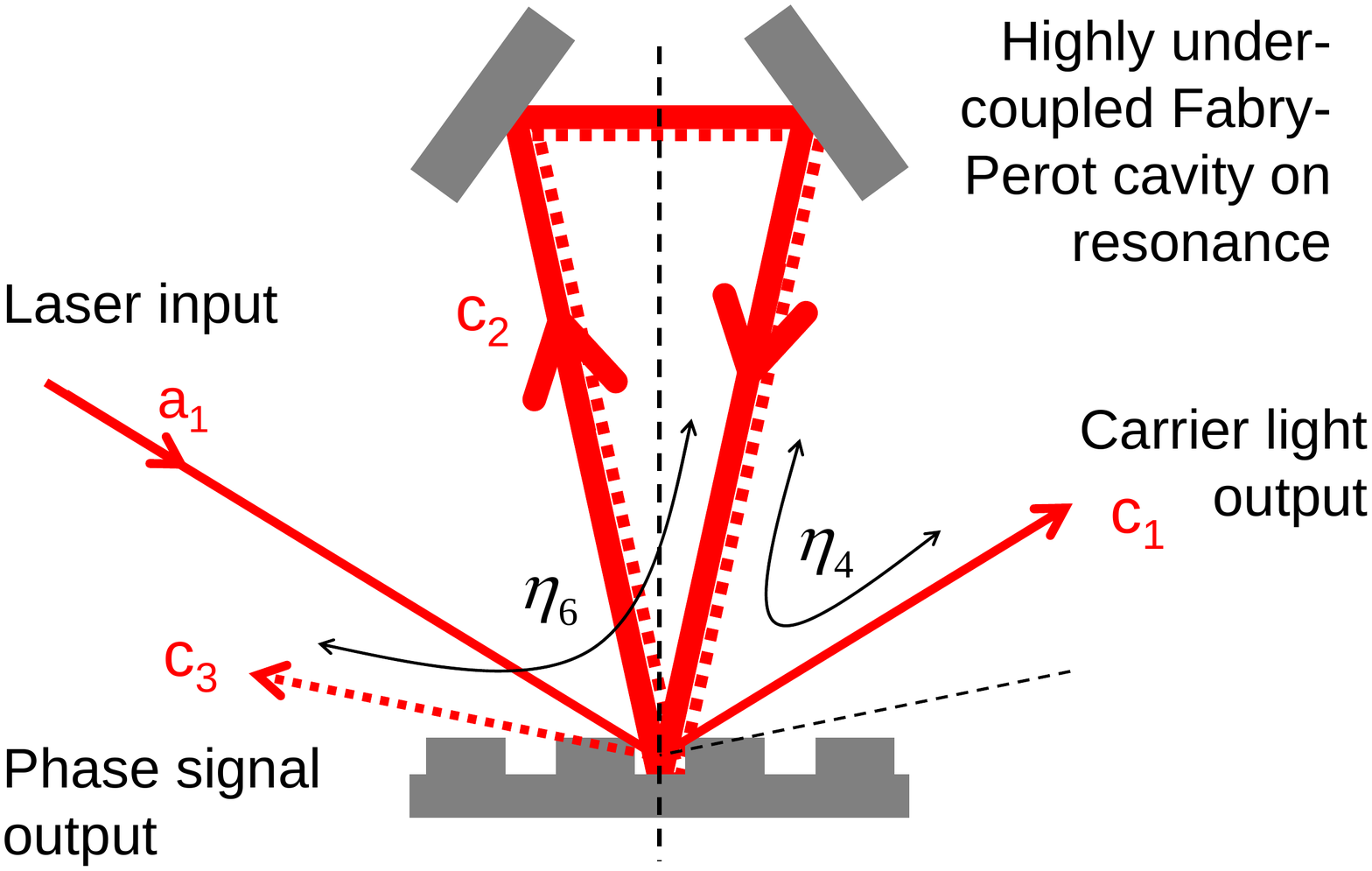}}
  \caption{Three-port-coupled grating in a ring Fabry-Perot interferometer. The grating can be designed such that the laser input is completely sent into port $c_{1}$ on cavity resonance. If the cavity is impedance matched this device might serve as an all-reflective mode-cleaner.
Another interesting case occurs in which the cavity is highly
under-coupled. Then almost the complete cavity strain signals are
sent to port  $c_{3}$. Such a device separates carrier light from
its modulation sidebands.}
  \label{3portgrat6FPRC}
\end{figure}

We now apply a 3p beam splitter/combiner in interferometry. We
focus on the device in Fig.~\ref{3portgrat6}b as a coupler to a
Fabry-Perot ring cavity as shown in Fig.~\ref{3portgrat6FPRC}.
Laser light incident from the left is
%weakly coupled ($\eta_{5}^2 \ll 1$)
coupled according to $\eta_{4}^2 $  into the cavity which is formed by the
grating and two additional highly reflecting cavity mirrors. If
both cavity mirrors are loss-less the cavity finesse depends on the
specular reflectivity $\eta_{2}^2$ and does not rely on high
values of first or second order diffraction efficiencies. Using
high reflection dielectric coatings high finesse values and high
laser buildups are possible similar to the linear cavity
investigated in Ref.\cite{BBBDSCKT04}. However, here the cavity outputs
depend on $\eta_{4}^2 $ (into port $c_{1}$) and $\eta_{6}^2
$ (into port $c_{3}$) that can have different values.

Assuming unity laser input and perfectly reflecting cavity mirrors
the system is described by
\begin{equation}
\label{io-relation} \left( \begin{array}{c}
        c_{1}   \\
        c_{2}   \\
        c_{3}   \\
       \end{array}
\right )
 \;=\;\textbf{S} \!\times\!
\left( \begin{array}{c}
        1   \\
        c_2 \,\mbox{exp}(2i\theta)   \\
        0
       \end{array}
\right )\,.
\end{equation}
Here $\theta=\omega L / c$ denotes the detuning from cavity resonance; with $L$ the cavity length, $\omega$ the laser field angular frequency and $c$ the speed of light.
Solving for
the reflected amplitudes yields
\begin{eqnarray}
\label{cavity}
\label{c1}
c_{1}&=& \eta_1 +\frac{\eta_4^2 \exp[2i(\phi_{4}+\theta)]}{1-\eta_2\exp(2i\theta)}\\
\label{c2}
c_{2}&=& \frac{\eta_4\exp(i\phi_{4})}{1-\eta_2\exp(2i\theta)}\\
\label{c3}
c_{3}&=& \eta_5\exp(i\phi_{5}) + \frac{\eta_4\eta_6\exp[i(\phi_{4}+\phi_{6}+2\theta)]}{1-\eta_2\exp(2i\theta)}~~~~
\end{eqnarray}

From Eq.~(\ref{c1}) it can be shown that for a grating with
$\eta_5^2$ at its maximum value for given $\eta_{4}^2$ and
$\eta_{6}^2$, and a cavity on resonance ($\theta=0$) no carrier
light from the laser incidenting from the left is leaving the
cavity to the left ($c_{3}=0$). This dark port is indicated in
Fig.~\ref{3portgrat6FPRC} by an arrowed dashed line. If the cavity
moves away from resonance for example caused by a cavity strain,
amplitude  $c_{3}$ is no longer zero. This field is generally
termed a phase signal and might appear at some sideband frequency
$\Omega$ if the cavity is locked to the time averaged carrier
frequency $\omega_{0}$ with locking bandwidth smaller than
$\Omega$. The phase signal generated inside the cavity obviously
leaves the cavity according to the magnitudes of  $\eta_{4}^2$ and
$\eta_{6}^2$ in two directions. From Eqs.~(\ref{c1}) and
(\ref{c3}) it is easy to prove that the power of the signal indeed
splits according to the ratio $\eta_{4}^{2}/\eta_{6}^{2}$.
We now discuss two distinct examples; in both of them we consider
$\eta_5^2$ to be designed close to its maximum value.
% practically a local oscillator is needed anyway
For
$\eta_{4}^{2} = \eta_{6}^{2}$ the cavity output coupling is twice
the input coupling and the signal is split into two equal halves.
We term this case a \emph{symmetric} or an \emph{impedance
matched} three-port coupled cavity; this is in analogy to the
loss-less impedance matched linear cavity whose output coupling is
also twice the input coupling. However, due to the choice of $\eta_5^2$ all the carrier power is
sent into port $c_{1}$ if the cavity is on resonance as discussed
above. Such a device can serve as an all-reflective
mode-cleaner. For $\eta_{4}\!>\!\eta_{6}$ the 3p coupled loss-less
cavity can be termed \emph{over-coupled} and for
$\eta_{4}\!<\!\eta_{6}$ \emph{under-coupled}.
As the second example we consider the highly under-coupled grating
cavity ($\eta_{4}^2 \ll \eta_{6}^2 \ll\eta_{2}^2$) and explicitly choose the following coupling coefficients
\begin{equation}\label{dataset}
\begin{array}{rlrlrl}
   \!\!\eta_4^2\!\!\!&= 0.0001\,,   &    \, \eta_{6}^2 \!\!\!&= 0.0099\,,     &  \, \eta_{2}^2   \!\!\!&= 0.99 \,, \\ [0.3mm]
   \!\!\eta_5^2\!\!\!&= 0.0394\,,   &    \, \eta_{1}^2  \!\!\!&= 0.9605\,,    &  \, \eta_{3}^2 \!\!\!&= 0.9507\,,\\
        \!\!  \phi_{1}\!\!\!&= 0\,,        &    \, \phi_{2}  \!\!\!&= 0\,,                  &  \, \phi_{3} \!\!\!&= 0                          \,,\\
  \!\!\phi_{4} \!\!\!&\approx -3.1349\,,   &     \,\phi_{5}     \!\!\!&\approx 1.5708\,,   &     \,\phi_{6} \!\!\!&\approx 1.5707\,.
\end{array}
\end{equation}
For this set of measures again $\eta_5^2$ is almost at its maximum value and consequently $\eta_{1}^2$ and $\eta_{3}^2$ are close to their minimum values.
As in the impedance matched case described above again all the
carrier power is sent into port $c_{1}$. Due to the high asymmetry
of the ratio between $\eta_{4}^2$ and $\eta_{6}^2$ the
\emph{signal} is mainly sent into port $c_{3}$. The special
property of the highly under-coupled grating Fabry-Perot
interferometer is therefore the possibility of separating carrier
light and phase signal. This is a remarkable result.
Separation of carrier light and phase signal is well known for a
Michelson interferometer operating on a dark fringe. Such an
interferometer sends all the laser power back to the laser source.
The antisymmetric mode of phase shifts in the Michelson arms is sent into the dark port. The symmetric mode is combined
with the reflected laser power and sent towards the bright port.
In case of the highly under-coupled 3p grating
Fabry-Perot interferometer the almost \emph{complete} phase signal
is separated from carrier light and is accessible to detection and
the reflected field in the bright port contains only a marginal
fraction of the signal ($\eta_{4}^{2}/\eta_{6}^{2}$).

We point out that all results obtained for the Fabry-Perot ring
interferometer using the 3p coupler in Fig.~\ref{3portgrat6}b
also hold for a linear cavity using the 3p coupler in
Fig.~\ref{3portgrat6}a. However, some distinctive properties
should be mentioned. Regardless of their different topologies the
ring FP-interferometer is content with only low efficiencies for
greater than zero diffraction orders. All coupling amplitudes in
Eqs.~(\ref{dataset}) with values close to unity describe specular
reflections. The production of such a grating with low overall
loss should be possible with standard technologies building on the
concept used in Refs.\cite{BBBDSCKT04,CKTBBDSDG05}. In case of the
(highly under-coupled) \emph{linear} FP-interferometer $\eta_{1}^2$ and
$\eta_{3}^2$ do not describe specular reflections and high
diffraction efficiencies in the second order diffraction is
required.
However, especially in the second order Littrow configuration carrier and signal separation offers the extension by interferometer recycling techniques \cite{HSMSWWSRD98}.
Recycling techniques in combination with a grating coupled Fabry-Perot
cavity will be subject to an upcoming publication \cite{SBBD05}.

This work was supported by the Deutsche Forschungsgemeinschaft
within the Sonderforschungsbereich TR7. R. Schnabel's email
address is roman.schnabel@aei.mpg.de.

\end{document}